# Equilibrium structure and shape of Ag and Pt nanoparticles grown on silica surfaces: from experimental investigations to the determination of a metal-silica potential

F. Ait Hellal[1], C. Andreazza-Vignolle[1], P. Andreazza[1], J. Puibasset[1*]

[1] ICMN, CNRS, Université d'Orléans, 1b rue de la Férollerie, CS 40059, 45071 cedex 02, Orléans, France

* corresponding author : puibasset@cnrs-orleans.fr

abbreviations :

nanoparticle (NP)

Surface Plasmon Resonance (SPR)

X-ray photoemission spectroscopy (XPS)

Density Functional Theory (DFT)

Grazing-incidence small-angle X-ray scattering (GISAXS)

Transmission Electron Microscopy (TEM)

High Resolution Transmission Electron Microscopy (HRTEM)

Tight Binding in the Second Moment Approximation (TBSMA)

truncated octahedron (TOh)

aspect ratio (AR)

Atomic Force Microscopy (AFM)

Distorted wave Born approximation (DWBA)

**ABSTRACT:**

A combination of experimental and numerical calculations on metallic silver and platinum nanoparticles deposited on silica substrates is presented, with a focus on the metal-substrate interactions. Experimentally, the nanoparticles are elaborated under ultra-high vacuum and characterized by Grazing-Incidence Small-Angle X-ray Scattering (GISAXS) and High Resolution Transmission Electronic Microscopy (HRTEM) to determine their structure and morphology, and in particular their aspect ratio (height/diameter) which quantifies the metal-substrate interaction. Numerically, the interactions between the metal and the silica species are modeled with the Lennard-Jones (12, 6) potential, with two parameters for each metal and silica species. The geometric parameters were found in the literature, while the energetic parameters were determined from our experimental measurements of the aspect ratio. The parameters are: $\sigma_{Ag-O}$ = 0.278 nm, $\sigma_{Ag-Si}$ = 0.329 nm, $\varepsilon_{Ag-O}$ = 75 meV, and $\varepsilon_{Ag-Si}$ = 13 meV for Ag-silica and $\sigma_{Pt-O}$ = 0.273 nm, $\sigma_{Pt-Si}$ = 0.324 nm, $\varepsilon_{Pt-O}$ = 110 meV, and $\varepsilon_{Pt-Si}$ = 18 meV for Pt-silica. The proposed Ag-silica potential reproduces quantitatively the unexpected experimental observation of the variation of the aspect ratio for Ag nanoparticles larger than 5 nm, which has been interpreted as a consequence of the silica roughness. The nanoparticle orientation, structure and disorder are also considered. This metal-silica potential for Ag and Pt should be helpful for further studies on pure metals as well as their alloys.

## 1-INTRODUCTION

Reducing the size of metallic particles up to the nanometer scale is the promise to reveal new properties or to exalt properties of the bulk or the surface of materials.[1,2] These properties that are dependent on the metal nature (confinement effect, surface effect), are also strongly influenced by the close environment of nanoparticles (NPs), including ligands, matrix or supports on which the NPs are elaborated or deposited.[3,4,5]

In the case of supported NPs, the features of the substrate influence their stability with modification of their morphology or structure.[6,7] It may also influence the NPs' physico-chemical properties, leading to improved efficiency for targeted applications. For example, the catalytic activity of a metallic NP not only depends on its size and morphology, but also on the choice of the support and its nature (amorphous *vs* crystalline) since charge transfer between the metal and the support can take place.[8,9,10,11]

Conversely, it may be desirable to minimize the influence of the environment by choosing a support in weak interaction with the NP metal atoms, like amorphous silica or carbon. This is a routinely used technique to approach the free NP (bare NP) situation in experiments dedicated to study the NPs' intrinsic properties. However, even in this case, it has been shown that the substrate can influence the morphology and structure of NPs.[12] So, further investigations including the influence of the support are required to correlate the experimental observations to theoretical models by a cross-fertilization.

Silver and platinum exhibit interesting catalytic properties that can be combined in Ag-Pt nanoalloys.[13,14] For example, Ag allows the dissociative adsorption of oxygen while Pt adsorbs and dissociates CO and C-H molecules.[15,16,17] AgPt could be a performant catalyst for CO oxidation, and few studies investigated their efficiency when supported on silica supports.[18,19] A better theoretical understanding of the influence of the substrate on the nanoalloy's properties would pave new paths to experimentalists.

In this context, reliable metal-silica potentials for molecular simulations of noble metals like silver and platinum on silica surfaces are requested. For the Ag-SiO$_2$ system, Eusthatopolous *et al.* have performed measurements of wetting angles of liquid silver droplets deposited on silica surfaces.[20] The large wetting angle (140°) is an indication of weak wettability on the substrate. Several evidences support the same conclusion: Tilted-view scanning electronic microscopy on truncated Ag nanospheres on quartz (001) revealed an aspect ratio (defined as the height of the supported NP over its diameter) of 0.85.[21] X-ray photo-emission spectroscopy (XPS) measurements have shown that the electronic configuration of an embedded silver atom in silica is equivalent to that of a gas phase free atom and that charge transfer is negligible.[22,23] From a theoretical point of view, the *ab initio* calculations performed by Vakula and co-workers on the (001) surface of α-quartz confirmed that the Ag-silica interaction is weak on model surfaces free of dangling bonds or defects.[24,25] They showed that the 1×1 reconstruction of the (001) surface of α-quartz exhibits two weakly attractive sites for isolated Ag atoms with adsorption energies of -60 and -80 meV.[24] Ngandjong *et al.*[26,27] performed Density Functional Theory (DFT) calculations of Ag atoms adsorbed on hydroxylated β-cristobalite, another silica polymorph, and found an adhesion energy of -40 meV, confirming that the metal-silica interaction is weak for hydroxylated surfaces, and proposed a van der Waals like potential to describe the interaction. On the other hand, surfaces presenting reactive defects (dangling oxygen, small rings, surface stress…) exhibit high adsorption energies.[28][29] Such situations are however not expected to occur in experiments when the silica surfaces are exposed to air and fully hydroxylated.

The Pt-SiO$_2$ system is less known. Levine and Garofalini considered the adsorption of Pt atoms and the growth of Pt NPs on silica supports using a Lennard-Jones potential.[30][31][32] However, their Pt-O potential actually corresponds to that of an isolated atomic O in contact with the Pt (111) surface, a highly reactive configuration.[33] Ewing *et al.* performed DFT calculations of small Pt$_{13}$ clusters on hydroxylated amorphous silica, and found a strong interaction between metal and silica, with charge transfer.[34][35] However, these results were not confirmed by Plessow *et al* who argued that the silica surfaces used by Ewing and collaborators were probably unstable and unoptimized.[36] The DFT calculations of Pt in contact with the (001) surface of α-quartz performed by Plessow *et al.* give adhesion energies for isolated Pt atoms of -220 meV and -810 meV on the reconstructed and hydroxylated surfaces respectively. For Pt slabs, the interaction is weaker and mainly due to the van der Waals contributions: -80 meV per atom in contact with both the reconstructed or the hydroxylated surfaces.[36] Experimental observations have shown that charge transfer can occur between Pt and silica when deposited in a vacuum or annealed above 250°C in air, with an aspect ratio of about 0.7.[23][37][38] These results confirm that the Pt-silica interaction is weak and most probably dominated by van der Waals contributions, similarly to the Ag-SiO$_2$ system.

Our objective is to propose a coherent parametrization for Ag and Pt interactions with silica surfaces in order to model Ag, Pt and AgPt NPs supported on silica substrates. Considering the sparse available data in the literature and the recognized influence of the structure and surface chemistry of the silica support, we have devised experiments to measure the aspect ratio of pure Ag and Pt NPs on silica supports to have access to the adhesion energies. We have carried out grazing-incidence small-angle X-ray scattering (GISAXS) on the supported NPs to obtain information about the morphology: size, height and aspect ratio. This approach to the metal-silica interaction has the advantage of taking into account the specific features of the support at a phenomenological level (hydroxylation, surface defects, SiO$_2$/Si thickness, etc.). However, it does not provide any information at the atomic scale, in particular regarding the relative importance of Si and O contributions to the van der Waals interactions, and the corresponding van der Waals diameters between the metal and silica species. Such information can be provided by theoretical studies found in the literature.[26][27][39]

The paper is structured as follows: first the samples elaboration procedure is described, as well as experimental and numerical methods to obtain the NPs aspect ratio when they are deposited on a silica amorphous support. Then the main part of this paper is dedicated to the results leading to the determination of the metal-silica potential. The originality of our approach is the combination of experimental and numerical studies to determine a robust potential allowing the exploration of a large range of size, structure and orientation. This potential is further used to propose an interpretation for the counter intuitive evolution of the nanoparticle morphology with size.

## 2 MATERIALS AND METHODS

### 2-1: Samples preparation and characterization

Ag and Pt metallic nanoparticles were prepared by atomic deposition under ultra-high vacuum conditions on two amorphous silica substrates of several mm² made of thermally oxidized Si(001) wafers and copper Transmission Electron Microscopy (TEM) grids coated by amorphous silica. Before metal's deposition, the substrates were annealed under ultra-high vacuum at 473 K for 30 minutes. This step is important to get rid of organic contamination and hydroxyl groups on the surfaces. The atomic depositions were carried out at room temperature, with a deposition rate in the same range of $0.73 \times 10^{15}$ at/cm²/hour for Ag and $0.6 \times 10^{15}$ at/cm²/hour for Pt (close to 0.5 atom monolayer/hour-ML/h). This protocol, performed at a substrate deposition temperature of 300K, is expected to produce Volmer-Weber mode nanoparticles with sizes of few nanometers on amorphous substrates.[40][41]

High Resolution Transmission Electron Microscopy (HRTEM) was used to investigate the morphology and the structure of the supported NPs with an in-plane view of the NP assembly, while GISAXS measurements were carried out to provide the in plane and out-of-plane morphology (average diameter and height and their relative distribution) and the NPs' organization on the substrate.[40][42] As shown on Fig. 1, from the 2D GISAXS pattern (a) of the reciprocal space, two cuts of intensity, i.e. 1D profiles, in the (b) $q_z$ and (c) $q_y$ directions were extracted and analyzed. Indeed, the analysis procedure consists of simultaneously fitting with a dedicated code these two perpendicular experimental cuts (marks) selected in the lobe intensity region (white lines in Fig. 1a). This IsGISAXS software is able using the distorted wave Born approximation (DWBA) framework and a truncated-sphere model as the NP shape to reveal their morphological features.[43][44][45]

For comparison between GISAXS and HRTEM measurements, the metal is deposited simultaneously on the two supports. The GISAXS chamber, kept at $10^{-5}$ mbar pressure and equipped with a furnace, allows the annealing of the samples during X-ray scattering data collection to reveal the possible morphological evolution from as grown metastable to equilibrium NP state. Samples were annealed from room temperature up to 800 K with a ramp of 5 K/min, and GISAXS acquisitions were performed every 100 K.

**2-2: Elaboration of the numerical substrates**

S. Tsuneyuki *et al.* developed a potential that reproduces the tetrahedral $SiO_4$ structure in silica with simple two-body spherically symmetric potentials acting on $Si^{4+}$ and $O^{2-}$ ions.[46] This potential of the Born-Huggins-Mayer analytical form takes into account the electrostatics, soft core, and dispersion forces. It has proven to be precise enough to describe the different crystalline silica polymorphs as well as the amorphous silica. However, at a very high temperature in the liquid phase, the atom-atom repulsion barrier at short distance is not high enough compared to the kinetic energy to prevent atomic collapse. Guissani and Guillot cured this problem by introducing a short-range repulsive term extending the validity of the potential up to 5000 K.[47]

The amorphous silica surface is produced as follows.[26][27] First, a piece of bulk silica melt is run at 2000 K to produce a disordered structure. It is then slowly cooled down to 1000 K where atomic mobility has become negligible (relaxation times larger than simulation runs). The surface is cut and relaxed to allow surface atomic reorganization. The creation of a free surface creates highly reactive dangling bonds like Si-O°, which have to be saturated with hydrogen to form silanols Si-O-H. This procedure mimics a realistic silica surface that has been put in contact with air. Note that the relaxation of the free surface before hydroxylation induces a low level of hydroxylation (1 OH/nm$^2$) compatible with the outgassing of the thermally oxidized wafers. The dimensions of the amorphous silica slab are 3.4×3.4 nm$^2$ (see Fig. 2). For large NPs, we have used 2×2 or 3×3 replications of this surface in the *x* and *y* directions to avoid cross-talk between the periodic images of the NP. In all cases, after the substrates have been elaborated, the silica species are frozen during the simulation runs with the metallic NPs to save computing time.

## 2-3: Metal-metal potentials

The metal-metal interaction was modeled with the semi-empirical many-body potential derived from the tight binding in the second moment approximation (TBSMA).[47][49] The energy $E_n$ at each site n is the sum of two contributions: $E_n = E_n^b + E_n^r$, where $E_n^b$ is an attractive many-body term from the band energy and $E_n^r$ is a core-repulsion term of the Born–Mayer type:

$$E_n^b = -\left\{\sum_{m \neq n} \xi_M^2 \exp\left[-2q_M \left(\frac{r_{mn}}{r_M^0} - 1\right)\right]\right\}^{1/2} \quad (1)$$

$$E_n^r = \sum_{m \neq n} A_M \exp\left[-p_M \left(\frac{r_{mn}}{r_M^0} - 1\right)\right] \quad (2)$$

where the sums run over the sites m ≠ n, M denotes the metal (Ag or Pt), and $r_{mn}$ is the distance between n and m. The summations are completed up to the second neighbors, and smoothly switched to zero between second and third neighbors with a fifth order polynomial to ensure continuity and derivability of the potentials. $\xi_M$, $A_M$, $q_M$, $p_M$ and $r_M^0$ are the potential parameters for pure crystal metals, fitted to the bulk cohesive energies, lattice parameters and elastic constants at zero temperature.[50][51][52][53][54][55]

## 2-4: Metal-silica potential

**General form.** As shown in the Introduction, the available data for the interaction between silver or platinum and a hydroxylated amorphous silica surface suggest that it is weak and can be approximated by a simple van der Waals form as given by the (12-6) Lennard-Jones potential:

$$V_{M-S}(r) = 4\varepsilon_{M-S}\left[\left(\frac{\sigma_{M-S}}{r}\right)^{12} - \left(\frac{\sigma_{M-S}}{r}\right)^6\right] \quad (3)$$

where $V_{M-S}$ is the interaction potential between the metal M (Ag or Pt) and the silica species S (Si, O), *r* is the distance between M and S, $\varepsilon_{M-S}$ is the well depth and $\sigma_{M-S}$ the distance at which the potential between atoms M and S is zero. Following Ngandjong *et al.* [26][27] we neglect the hydrogen species in the description of the interaction; however, their energetic contribution is taken into account at a phenomenological level into the Si and O contributions. For each metal M, one is left with four parameters $\sigma_{M-Si}$, $\sigma_{M-O}$, $\varepsilon_{M-Si}$ and $\varepsilon_{M-O}$ that have to be determined.

The geometric parameters are available in the literature for the pure metallic species Ag and Pt ($\sigma_{Ag-Ag}$ = 0.2574 nm, $\sigma_{Pt-Pt}$ = 0.2471nm) as well as for the Ag-SiO$_2$ system. [26][27][39] For Pt-SiO$_2$ we use the Lorentz-Berthelot combination rules[56] to get $\sigma_{Pt-S}$ = ½ ($\sigma_{Pt-Pt}$ − $\sigma_{Ag-Ag}$) + $\sigma_{Ag-S}$, where S belongs to the silica species (Si and O). The values are given in Table 1.

For energetic parameters, Ngandjong *et al.*[26][27] have shown that the Ag-silica interaction is essentially dominated by the most polarizable silica species, *i.e.* oxygen. The same behavior has been reported for the Pt-silica system.[34] However, the Si species cannot be simply ignored because their repulsive contribution is important to prevent metal diffusion into silica. Ngandjong *et al.* have shown that $\varepsilon_{Ag-O}/\varepsilon_{Ag-Si} = 6$ for quartz, and keep the same ratio for the other silica substrates (cristobalite and amorphous $SiO_2$). We adopt the same procedure, keeping the ratio $\varepsilon_{M-O}/\varepsilon_{M-Si} = 6$ for Ag as well as for Pt. This way, we are left with only one free parameter for each metal. By convention, we choose the dominant $\varepsilon_{M-O}$ to be this free parameter.

**Procedure to determine the $\varepsilon_{M-O}$ parameter.** As previously shown, the metal-silica potential is almost entirely determined except for one energetic parameter $\varepsilon_{M-O}$. The latter can be easily deduced from the aspect ratio or equivalently the adhesion energy measurements on the silica substrate. The procedure consists in scanning different values for the free parameter $\varepsilon_{M-O}$ until the model reproduces the experimentally observed aspect ratio of the supported NPs.

The experiments being performed at room or moderate temperature (below the bulk melting point), the metal remains in the solid state and the aspect ratio measurements can safely be extrapolated to low temperatures. Therefore, the calculations are done at zero Kelvin, and the stable structures are those with the lowest energy. In our experiments, the size of the NPs, controlled by the amount of metal deposited, is in the range of a few nanometers, depending on the metal and the preparation parameters (substrate temperature). It was thus decided to perform calculations with NPs of few nanometers in diameter.

Free NPs are expected to adopt an almost spherical shape that minimizes their surface energy. When deposited on a support, their aspect ratio is lower than one due to the metal-support interaction. In this work, the aspect ratio is defined as the ratio between the height of the supported NP and its height before the effect of truncation due to the support, which also corresponds to its diameter for a spherical structure. The advantage of this definition is that the aspect ratio equals exactly one for free NPs, even for Wulff polyhedra. The atomic configurations of supported NPs are built by removing one by one the atomic layers in contact with the substrate (see Fig. 3), and their energy minimized with quenched Molecular Dynamics using the metal-metal TBSMA potential and the metal-silica potential with different values of $\varepsilon_{M-O}$.

An important point to be discussed concerns the effect of the amorphous silica surface disorder on the precision of the simulation results. The total NP-silica interaction strongly depends on the interatomic distances between the atoms of the NP close to the NP-support interface and the silica atoms at the surface. Obviously, for a rough and heterogeneous surface, the interaction will depend noticeably on the exact position on the surface. Figure 4 gives the distribution of the adhesion energies of a 4 nm Ag NP deposited on the amorphous silica substrate, for two aspect ratio corresponding to the removal of three and four atomic layers. As can be seen, the distributions overlap, meaning that the difference between two positions may be as large as for two successive aspect ratio: it is then required to perform an average on a set of surface positions to minimize the uncertainties. We have chosen nine positions uniformly distributed on the surface materialized as crosses in Fig. 2.

To discriminate which aspect ratio is the most favorable, it is sufficient to compare the average minimized energies (zero temperature calculations). The values are however dominated by the strongly cohesive metal contribution, and the number of atoms depends on the aspect ratio. This is why we introduce the excess energy Δ according to[57][58]:

$$\Delta = \left(E_{tot} - E_{SiO_2} - N e_{coh}^M\right)/N^{2/3} \qquad (4)$$

where $E_{tot}$ is the energy of the total NP + silica system, $E_{SiO_2}$ is the energy of the silica substrate, $e_{coh}^M$ is the cohesive energy of the metal M ($e_{coh}^{Ag}$ = -2.95 eV and $e_{coh}^{Pt}$ = -5.86 eV), $N$ is the total number of atoms in the NP, and $N^{2/3}$ is proportional to the number of surface atoms of the NP. In other words, we compare our systems to a piece of bulk metal of the same size, normalized to the number of surface atoms.

## 3-Results and discussion

### 3-1: Aspect ratio of the supported NPs on silica

The first sample to be considered is Ag deposited at room temperature on amorphous silica, the total amount being 2.9×10$^{15}$ at/cm$^2$. HRTEM observations show that the NPs are essentially spherical and isolated. Their size distribution is centered around 3.9 nm ± 1 nm (see Fig. 5). After annealing at 573 K for 4 hours, one observes a slight increase of the average size to 4.1 nm. HRTEM observations also show that the structure of the NPs is affected by the annealing. As depicted in Fig. 6, as deposited Ag NPs exhibit crystalline as well as non-crystalline structures in large proportion (decahedron and icosahedron), while after annealing the crystalline FCC structure dominates. This motivates the choice of the truncated octahedron (TOh) structure to perform the atomic simulations.

This increase in size during annealing is corroborated by GISAXS measurements, which make it possible to monitor changes in diameter and height during the various annealing stages. The evolution of the aspect ratio during the annealing, deduced from GISAXS measurements is shown in Fig. 7. The aspect ratio slightly decreases above 343 K from 0.81 to 0.78, which may be correlated to the dehydroxylation of the silica substrate, in analogy with Pb on quartz substrate.[59] Beyond this temperature, the aspect ratio has essentially a constant value equal to 0.78 for NPs with an average size of 4.1 nm, that has been used to determine the Ag-silica potential.

A similar study has been done for Pt NPs. Unlike the Ag case, Pt NPs branch out very quickly. For a deposited quantity of $1.8 \times 10^{15}$ at/cm², the Pt nanoparticle density is three times greater than that observed for the Ag sample described above. Nanoparticles are not isolated and exhibit noticeable ramifications (static coalescence during deposition),[40] which are expected to evolve during annealing, with more spherical and less branched shapes. HRTEM observations reveal only crystalline structures. The average diameter of Pt NPs increases from 1.6 nm to 1.9 nm during annealing, and their aspect ratio, given by GISAXS measurements, increases from 0.72 to 0.84 as shown in Fig. 7. The aspect ratio obtained after annealing most probably represents the equilibrium value for Pt NPs of average size 1.9 nm, and has been used to determine the Pt-silica potential.

### 3-2: Determination of an interval for $\varepsilon_{M-O}$

The excess energy for the free and supported Ag and Pt NPs are given in Fig. 8 as a function of their aspect ratio and for an arbitrarily chosen value of the free parameter $\varepsilon_{M-O}$ ($\varepsilon_{Ag-O}$ = 72 meV and $\varepsilon_{Pt-O}$ = 156 meV). The NPs are deposited on the amorphous silica substrate with the denser (111) facet in contact with the support which favors the NP-support interaction. As can be seen, the excess energy is positive for all systems, showing that they are less stable than the corresponding bulk metals, as expected. In both cases, the free NP is the highest, meaning that the interaction of the NP with the substrate is favorable (van der Waals attraction). More interestingly, decreasing the aspect ratio gives lower excess energies. This is because removing the NP's bottom layer in contact with the silica results in a net increase in the number of metallic atoms in contact with the substrate, with a negligible penalty for the cohesive energy of the NP. For even lower aspect ratio, the energetic cost for the NP becomes larger because its shape departs significantly from the sphere: the total excess energy for the system becomes non-monotonic and exhibits a minimum. This minimum corresponds to the most stable aspect ratio for the supported NP, and is expected to be representative of the experimental measurements at room temperature (solid phase). In the examples given in Fig. 8, the most stable aspect ratio (AR) for Ag with $\varepsilon_{Ag-O}$ = 72 meV is AR = 0.78, and for Pt with $\varepsilon_{Pt-O}$ = 156 meV, AR = 0.74.

The same operation is repeated for different values of the free parameter $\varepsilon_{M-O}$ (which controls the intensity of the metal-silica interaction). The result is shown in Fig. 9. For increasing metal-silica interaction, the aspect ratio (AR) decreases, with an essentially linear dependence, as expected in the framework of a truncated sphere model or Wulff construction:

$$AR = 1 - \frac{E_{adh}}{2\gamma} \qquad (5)$$

where $E_{adh}$ is the adhesion energy per unit surface and $\gamma$ the surface tension. A linear fit of the data gives an effective surface tension equal to 0.30 eV/atom for silver NPs and 0.50 eV/atom for Pt NPs. These values are expected to be a combination of the free energies of the (111) and (100) facets of a TOh NP, in proportion with the total surface area of each facet. For Ag, $\gamma_{111}$ = 0.264 eV/atom and $\gamma_{100}$ = 0.340 eV/atom and, for, Pt $\gamma_{111}$ = 0.408 eV/atom and $\gamma_{100}$ = 0.549 eV/atom for Pt.[60] As can be seen, the $\gamma$ values are compatible with such a combination, albeit a precise comparison is plagued by the unknown contribution of the edges and vertices.

The vertical bars in Fig. 9 represent the uncertainties in the aspect ratio values, which have two origins. The first is that the aspect ratio of a small NP is necessarily discretized because of its definition in terms of layers (Fig. 3). The second comes from the fact that the localization of the minimum may reveal situations where two adjacent aspect ratio have essentially the same excess energy, as exhibited by Pt in Fig. 8. In that particular case, it is obvious that the two aspect ratio 0.74 and 0.79 are almost equally probable in terms of stability. One could define an average AR by affecting to each possible AR a Boltzmann weight deduced from its relative excess energy. Another possibility is to observe that the relative position of the points around the minimum vary continuously and smoothly with the free parameter $\varepsilon_{M-O}$. This is shown in the inset of Fig. 10 (upper panel) where the excess energy of a Ag NP of AR 0.78 is drawn as a function of $\varepsilon_{Ag-O}$. As can be seen, the excess energy decreases with increasing metal-silica interaction. The variations are essentially linear for $\varepsilon_{Ag-O}$ < 0.1 eV, as expected as long as the metal-support interaction remains negligible compared to the internal cohesion of the NP. Conversely, in the strong interaction regime, the deformation of the NP induces non-linear deviations.

The main panels of Fig. 10 display the excess energy of the Ag and Pt NPs as a function of the free parameter $\varepsilon_{M-O}$ for different AR around the experimental values: 0.84, 0.78 and 0.75 for Ag, corresponding to the removal of 3, 4 and 5 layers at the bottom of a 4.2 nm NP; 0.93, 0.86 and 0.79 for Pt, corresponding to the removal of 1, 2 and 3 layers at the bottom of a 3.2 nm NP. Note that the slope of the curves is related to the number of atoms at the interface. The lower the AR, the larger the metal-support interface, the lower the negative slope. For each value of the interaction parameter, the most stable AR is given by the lowest curve. The experimental AR are thus reproduced for $\varepsilon_{Ag-O}$ between 65 and 87 meV and $\varepsilon_{Pt-O}$ between 77 and 130 meV, with a better accuracy than in Fig. 9. More precise values will be proposed in next section.

### 3-3: Aspect ratio versus nanoparticle size

**Experimental results**. Size effects are expected to occur in nanometric systems. We have thus considered different sizes of supported Ag NPs on silica, which is controlled through the total amount of deposited Ag. Keeping a constant deposition rate of $0.73 \times 10^{15}$ at/cm$^2$/h, five samples have been prepared with a total deposition duration of 1h, 2.5h, 4h, 6h and 8h. After annealing, their measured average sizes are respectively 2.6, 3.4, 4.8, 5.9 and 6.8 nm and the corresponding aspect ratio are 0.77, 0.78, 0.78, 0.82 and 0.85. The results are shown in Fig. 11 (solid squares in the upper panel). The same procedure has been adopted for Pt. However, the elaborated Pt NPs present high ramification and non-spherical shapes when the amount deposited is increased. It was thus impossible to determine the AR for NP size larger than 2.5 nm (see Fig. 11, solid squares in the lower panel).

As can be seen, the values are essentially independent of the system size below 5 nm for Ag and below 2.5 nm for Pt, in agreement with the Wulff construction. However, for Ag, one observes a variation between 5 and 7 nm: such a variation is unexpected in the framework of a continuous drop model on a perfectly smooth substrate, but this model is helpful to interpret the data. Let us denote *D* the diameter of the drop, *n* the corresponding total number of layers, *d* the interlayer distance ($D = n\,d$), and *i* the number of layers removed from the bottom of the NP in contact with the support. The aspect ratio reads AR = 1-*i d/D*. The corresponding curves are given in Fig. 10 for Ag and Pt (solid lines) with the values of *i* given between parentheses on the right-hand side of the panel. The experimental points are not expected to fall exactly on the curves because the NP size is averaged on a distribution. The interesting point is that above 5 nm the data fall between $i$ = 4 and 5 while following the general trend of the theoretical curves, suggesting that the NPs between 5 and 7 nm actually keep a constant truncation (number of removed layers at the interface) instead of keeping a constant aspect ratio.

This variation of the aspect ratio reveals a dependence of the NP-support interaction with the NP size in the case of Ag. This paradoxical behavior may be explained by invoking the roughness of the amorphous silica substrate. For the smallest NPs, able to accommodate the local curvature of small hollows, the NP-support interaction is the strongest. On the other hand, NPs larger than the typical roughness of the support are constrained to remain on top of the hills, which reduces the effective NP-support interaction, and increases their AR. Experimental measurements on the silica support by Atomic Force Microscopy (AFM) reveal an average roughness of order 1 nm. This measure is however only indicative, since NPs will preferentially nucleate in the most attractive sites which may correspond to larger values.

**Simulation results**. What can be learned from numerical simulations? We have considered different sizes of NPs going from 2 to 6.5 nm, supported on the amorphous silica surface. We have considered different values of the interaction parameter $\varepsilon_{M-O}$ in the intervals previously determined. For each NP size, the excess energy has been calculated according to Eq. 4 and the most stable aspect ratio determined. The results are gathered in Fig. 11 for Ag and Pt (empty squares). The error bars correspond to the uncertainty in the determination of the aspect ratio due to the discretization and the relative positions around the minimum. Silver and platinum exhibit different behaviors. For Pt and $\varepsilon_{Pt-O}$ = 108 meV, Fig. 11 shows that the aspect ratio exhibits small erratic variations between 0.85 and 0.90 for a NP size between 1.8 and 6.0 nm. These variations actually originate in the discretization of the aspect ratio which has to fall on the closest theoretical curve for Pt ($d$ = 0.226 nm) compatible with a constant aspect ratio. A similar behavior is observed for silver and $\varepsilon_{Ag-O}$ = 84 meV: the aspect ratio of the NP in contact with the substrate is essentially independent of its size (AR = 0.78). Conversely, for the lowest value $\varepsilon_{Ag-O}$ = 72 meV, one observes a difference for the largest NP (6.5 nm) whose aspect ratio departs significantly (beyond discretization issue) from 0.78: the most stable truncation corresponds to the removal of 4 layers (AR = 0.85) instead of 6 for the expected AR = 0.78. The intermediate value $\varepsilon_{Ag-O}$ = 77 meV follows a similar behavior with a constant AR = 0.78 below 5 nm, and a noticeable increase for the largest NP (6.5 nm) with an AR = 0.82 corresponding to the removal of 5 layers.

**Model improvement**. Interestingly, the experimental behavior can be quantitatively reproduced for $\varepsilon_{Ag-O}$ between 72 and 77 meV. This may be used to improve the determination of the metal-silica potential. As a general trend, the discretization of the aspect ratio for NPs smaller than 5 nm induces large uncertainties in the determination of $\varepsilon_{M-O}$, while for larger NPs the adopted aspect ratio is increasingly more sensitive to the metal-silica interaction. However, from an experimental point of view, it is difficult to elaborate Ag NPs larger than 7 nm and Pt NPs larger than 2 nm (Pt NPs are particularly sensitive to coalescence and ramification). Gathering all data, we propose to use $\varepsilon_{Ag-O}$ = 75 meV and $\varepsilon_{Pt-O}$ = 110 meV as optimal values to describe the metal silica interactions. However, in the following section we continue to use $\varepsilon_{Ag-O}$ = 77 meV to allow quantitative comparisons with previous calculations.

### 3-4: Orientation, structure and disorder effect

**Orientation:** TOh NPs present (111) and (100) facets. Two orientations are thus possible for a deposited NP. We have determined the most stable aspect ratio for TOh NPs of size between 2 and 5 nm deposited on the (100) facet on the amorphous silica support. For $\varepsilon_{Ag-O}$ = 77 meV, the aspect ratio is constant on the whole size range and equals 0.82, a value slightly larger than for the NPs of comparable size and oriented with the (111) facet in contact with the substrate (see Fig. 12). A larger aspect ratio means a lower NP-support adhesion energy: the corresponding (100) orientation is thus expected to be less stable than the (111) orientation. This is why we have chosen the (111) orientation to determine the interaction parameter $\varepsilon_{Ag-O}$. In the case of Pt, we do not observe a different aspect ratio between the (100) and (111) orientations for TOh between 3 and 5 nm. The difference in adhesion energy for the two facets is less than the cost between two successive aspect ratio, possibly due to a higher cohesive energy of Pt.

**Structure:** HRTEM observations revealed the presence of some non-crystalline structures like icosahedron (Ih) and decahedron (Dh), as observed for Ag NPs on amorphous carbon by Nelli and al.[42]. Following the same procedure as before, we have determined the most stable aspect ratio for Ih on (111) facets and Dh on (111) and (100) facets and sizes varying between 1.5 and 5.5 nm. The results are given in Fig. 12. All structures exhibit aspect ratio above that of the TOh on the (111) facet. The largest value is for the Dh structures deposited on the (111) facet, while Ih on (111) and Dh on (100) facets have intermediate values. The TOh with (111) orientation thus appears to have the lowest aspect ratio and the highest adhesion energy which probably explains their relative abundance compared to the Ih and Dh structures. Similar results were obtained for Pt NPs. This behavior is probably quite general and can be explained by the fact that the truncation of a non-crystalline structure like Ih or Dh has a larger cost.

**Disorder:** Another question is the influence of internal disorder on the aspect ratio of the NP. When prepared by deposition at low temperature, the NPs may accumulate structural defects. At higher temperatures, the metal mobility is large enough to allow reorganization of the NPs (annealing), but with possibly local structure defects that may persist after annealing. We thus consider amorphous metallic NPs as a model to study the effect of defects. We melt a bulk Ag solid at 2000 K. It is then slowly cooled down until it forms an amorphous solid. We cut a sphere of 4.2 nm in diameter, which is then relaxed. We cut the bottom at different heights (aspect ratio between 1 and 0.7). The NP is deposited on the amorphous silica substrate, and energy minimized with $\varepsilon_{Ag-O}$ = 77 meV. The most stable structure is obtained for AR = 0.75. As can be seen in Fig. 12, this value is slightly lower than the 0.78 found for the TOh. This model suggests that the presence of disorder slightly decreases the aspect ratio. A possible interpretation is that the internal defects decrease the cohesive energy of the NP compared to that of the crystalline TOh, which results in a lower surface energy of the external facets, without affecting the NP-support interaction. It is however outlined that the effect is quite small for a fully disordered structure, and is possibly irrelevant for structures presenting few defects. The choice of a perfect crystalline structure for the determination of the metal-support interaction parameter is thus relevant.

## 4-CONCLUSION

A dual experimental and numerical analysis on Ag and Pt NPs was carried out to determine the effects of an amorphous silica support on the aspect ratio of the nanoparticles. Combined and complementary analyzes by HRTEM and GISAXS on nanoparticles of a few nanometers, in different conditions (as deposited or after annealing) made it possible to determine the predominant structure and aspect ratio. From these experimental data, it seemed relevant to use a Lennard-Jones potential to describe the metal-silica interaction. The obtained parameters are: $\sigma_{Ag-O}$ = 0.278 nm, $\sigma_{Ag-Si}$ = 0.329 nm, $\varepsilon_{Ag-O}$ = 75 meV, and $\varepsilon_{Ag-Si}$ = 13 meV for Ag-silica and $\sigma_{Pt-O}$ = 0.273 nm, $\sigma_{Pt-Si}$ = 0.324 nm, $\varepsilon_{Pt-O}$ = 110 meV, and $\varepsilon_{Pt-Si}$ = 18 meV for Pt-silica.

This potential allowed us to reproduce the paradoxical variation of the aspect ratio with the Ag NP size, and interpreted this variation as due to the substrate roughness (the small NPs can accommodate the local curvature of small hollows, while large NPs are constrained to remain on the top of the hills, resulting in a reduced effective interaction and a larger aspect ratio). We have also shown that the crystalline TOh structures present lower aspect ratio than the non-crystalline Ih and Dh structures of similar size, and that the NPs deposited on the (111) facet are more favorable, with a larger adhesion energy and a lower aspect ratio. The presence of defects in the NP has been modeled with a fully amorphous structure, which has revealed a weak effect on the aspect ratio of the deposited NP.

An interesting application of this potential is for the study of AgPt nanoalloys deposited on amorphous silica substrates, with the reasonable hypothesis that the metal-silica interaction remains additive for the alloy.[61]

## ACKNOWLEDGMENTS

F. A. H. acknowledges a grant from the education and research ministry for her Ph.D.

## AUTHORS DECLARATIONS

### Conflict of Interest

The authors have no conflicts to disclose.

### Author contributions

**F. Ait Hellal** : Data curation, Writing – original draft. **C. Andreazza-Vignolle** : Data curation, Writing – review & editing. **P. Andreazza** : Data curation, Writing – review & editing. **J. Puibasset**: Data curation, Writing – review & editing.

## DATA AVAILABILITY

The data that support the findings of this study are available from the corresponding author upon reasonable request.

**Table 1**: The Lennard-Jones geometric parameters $\sigma_{M\text{-}S}$ (nm) between metal M (Ag, Pt) and silica species S (Si, O) as given by refs [57] [31] [32] and deduced from combination rules (see text).[58]

|    | Si    | O     |
|----|-------|-------|
| **Ag** | 0.329 | 0.278 |
| **Pt** | 0.323 | 0.272 |

Figure 1

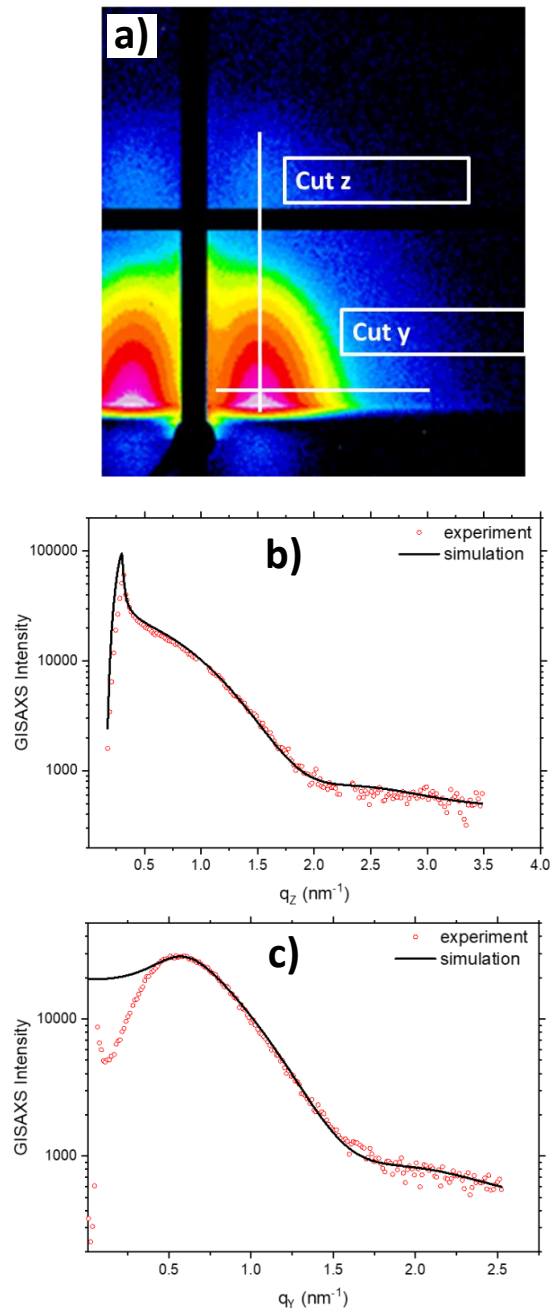

Fig. 1: a) Representative 2D GISAXS pattern of Ag NPs supported on amorphous silica. The $q_Y$ (resp. $q_z$) axis is parallel (resp. perpendicular) to the sample plane. b) and c) represent the 1D experimental measurements and the corresponding simulated fits for $q_z$ and $q_Y$ cuts respectively (shown by the white lines on the 2D image).

Figure 2

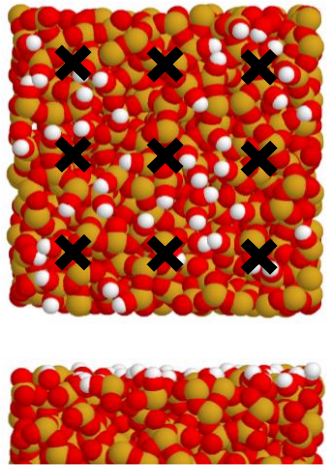

Fig. 2: Top (upper panel) and side (lower panel) views of the amorphous silica as obtained by quenching and cutting a piece of bulk liquid (see text). The horizontal dimension is 3.4 nm. Yellow: Si; red: O; white: H. The crosses materialize the nine uniformly distributed positions of the NPs used to perform the average on the surface disorder (see text).

Figure 3

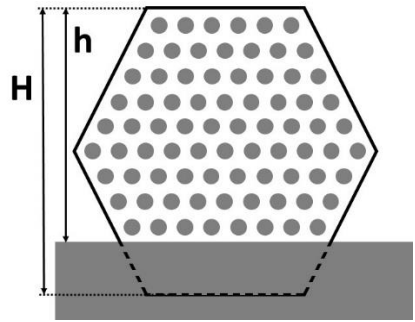

Fig. 3: Schematic lateral view of a NP (circles and solid line) deposited on a substrate (greyed region). The circles represent the projection of the aligned atoms. The solid line represents the envelope of the NP taking into account the van der Waals radius. The bottom of the NP, which is truncated due to its interaction with the support, is materialized as a dotted line. The aspect ratio (AR) is defined as AR = h/H, where h is the height of the truncated NP in interaction with the support, and H the height of the corresponding non-truncated free NP. With this definition AR = 1 when there is no interaction with the support. Typically, AR values are fractions ($n-i$)/$n$ where $n$ is the number of layers of the complete NP, and $i$ is the number of missing layers at the bottom.

Figure 4

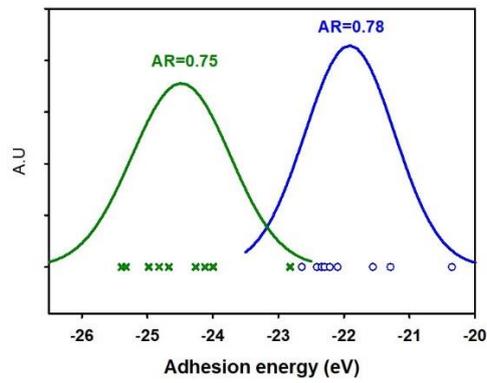

Fig. 4: Symbols: total adhesion energies (horizontal axis) of a 4 nm NP truncated at two consecutive aspect ratios AR (circles: AR = 0.78; crosses AR = 0.75; corresponding to the removal of 4 and 5 atomic layers respectively) in interaction with the amorphous silica substrate for the 9 positions depicted in Fig. 2. Vertical axis and lines: Gaussian distributions corresponding to the two data sets as given by the average and variance.

Figure 5

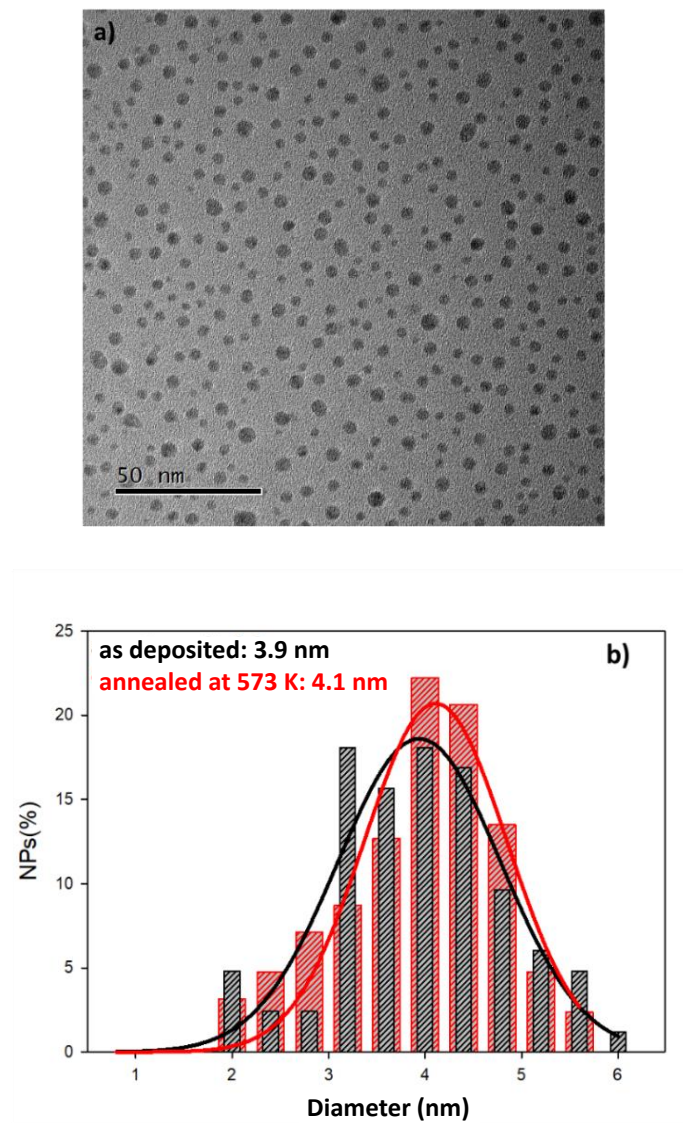

Fig. 5: a) TEM image of Ag NPs elaborated at room temperature on a silica coated copper grid with a total amount of 2.9×10$^{15}$ at/cm$^2$, and annealed at 573 K for 4 hours. b) Distribution of the NP diameters (histogram) with the corresponding Gaussian fit (line), before and after annealing.

Figure 6

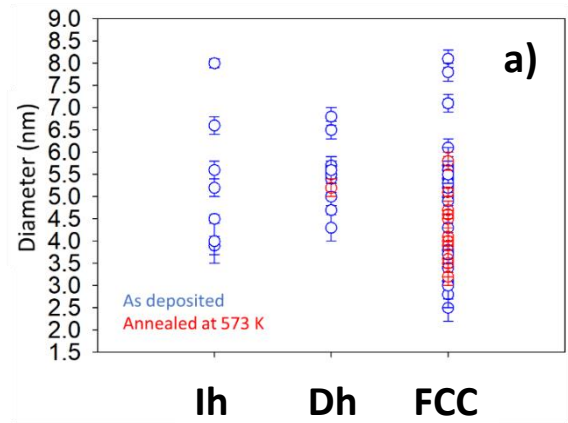
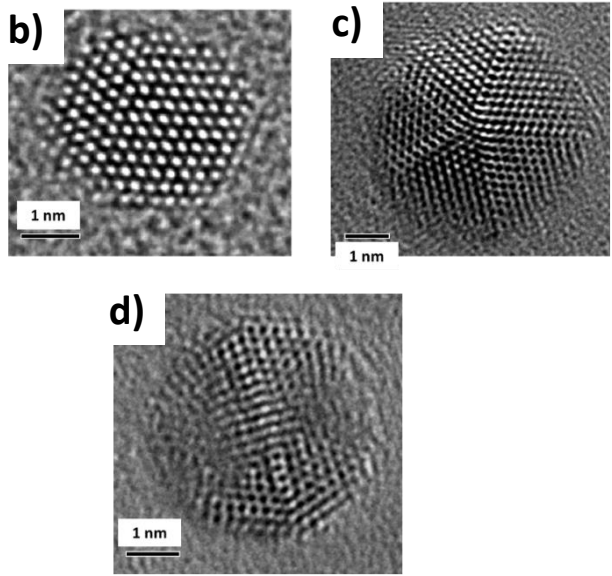

Fig. 6: a) Structure of silver NPs deposited on amorphous silica (a-SiO$_2$/Cu grids): as deposited (blue) and after annealing at 573 K for 4 hours (red). b) Crystalline FCC structure (3 nm), c) non-crystalline Dh (decahedron) structure (6 nm) and d) non-crystalline Ih (icosahedron) structure (5 nm).

Figure 7

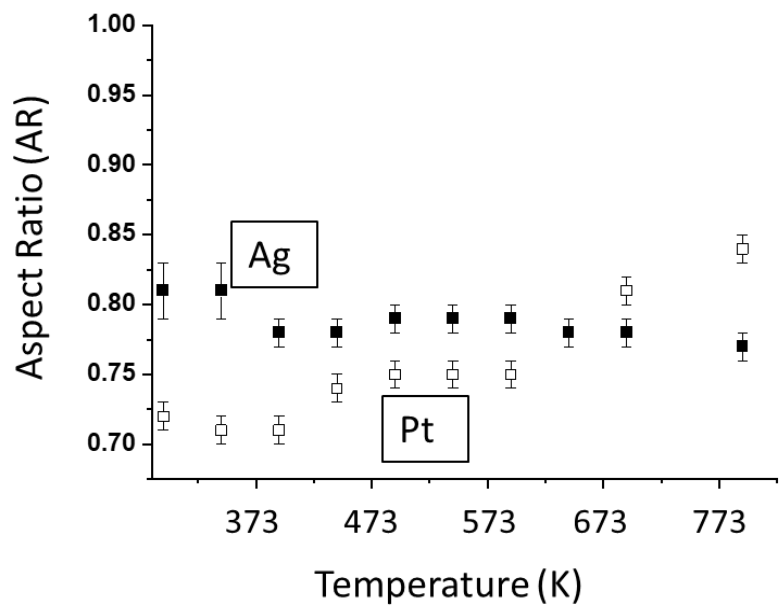

Fig. 7: Evolution of the aspect ratio deduced from GISAXS measurements of Ag and Pt NPs on amorphous silica during annealing as a function of the temperature. Filled symbols: Ag NPs ($2.9\times10^{15}$ at/cm$^2$); empty symbols: Pt NPs ($1.8\times10^{15}$ at/cm$^2$).

Figure 8

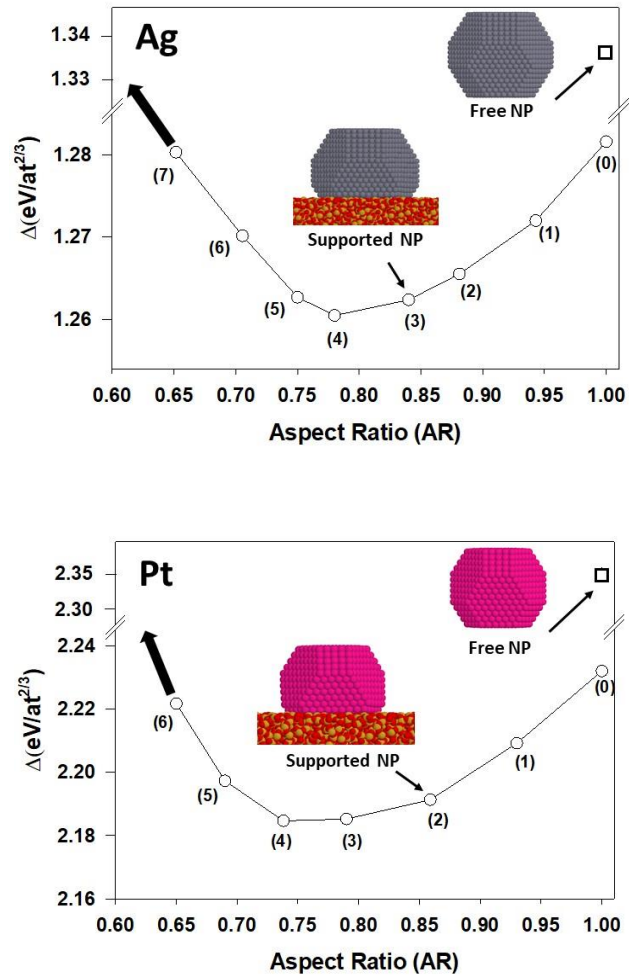

Fig 8: Excess energy Δ (Eq. 4) of free (square) and supported (circles) NPs as a function of their aspect ratio (defined in Fig. 3). For each aspect ratio, the corresponding number of removed atomic layers is given between parentheses. The supported NPs are oriented with the denser (111) facet in contact with the amorphous silica substrate. Upper panel: Ag NP of size 4.2 nm (TOh with 3348 atoms) and $\varepsilon_{Ag-O}$ = 72 meV; lower panel: Pt NP of size 3.2 nm (TOh with 1654 atoms) and $\varepsilon_{Pt-O}$ = 156 meV.

Figure 9

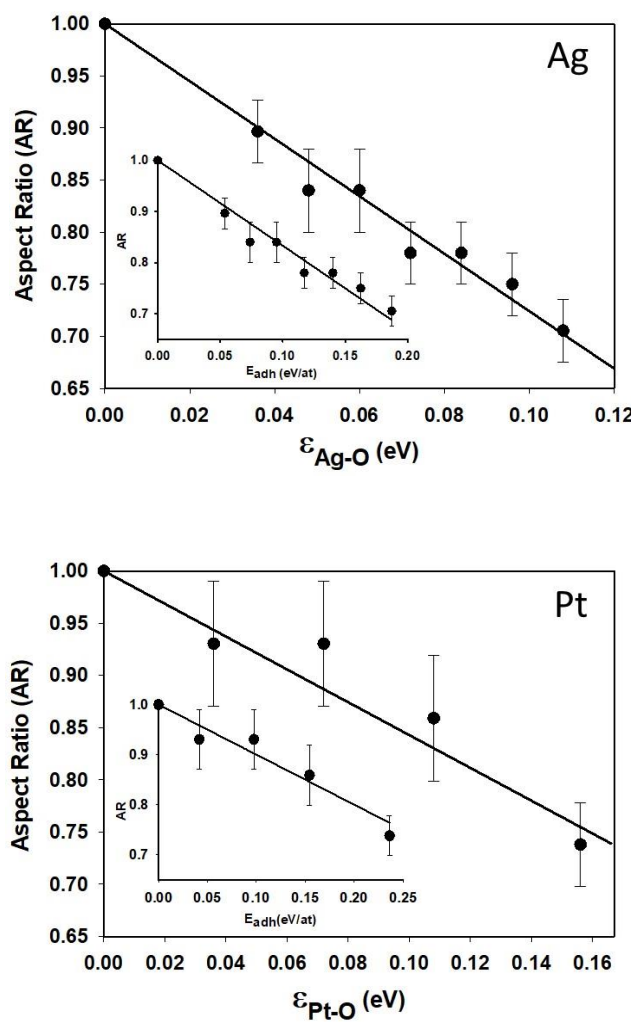

Fig 9: Aspect ratio for Ag (upper panel) and Pt (lower panel) NPs deposited on amorphous silica (circles) as a function of the free parameter $\varepsilon_{M\text{-}O}$ (main panels) or as a function of the adhesion energy $E_{adh}$ (inserts). The solid lines are linear fits through the data with the constraint to start from the upper left corner (AR = 1 for the free NP). The slopes of the AR vs $E_{adh}$ fits give the effective surface energies (Eq. 5): 0.30 eV/atom for Ag and 0.50 eV/atom for Pt.

Figure 10

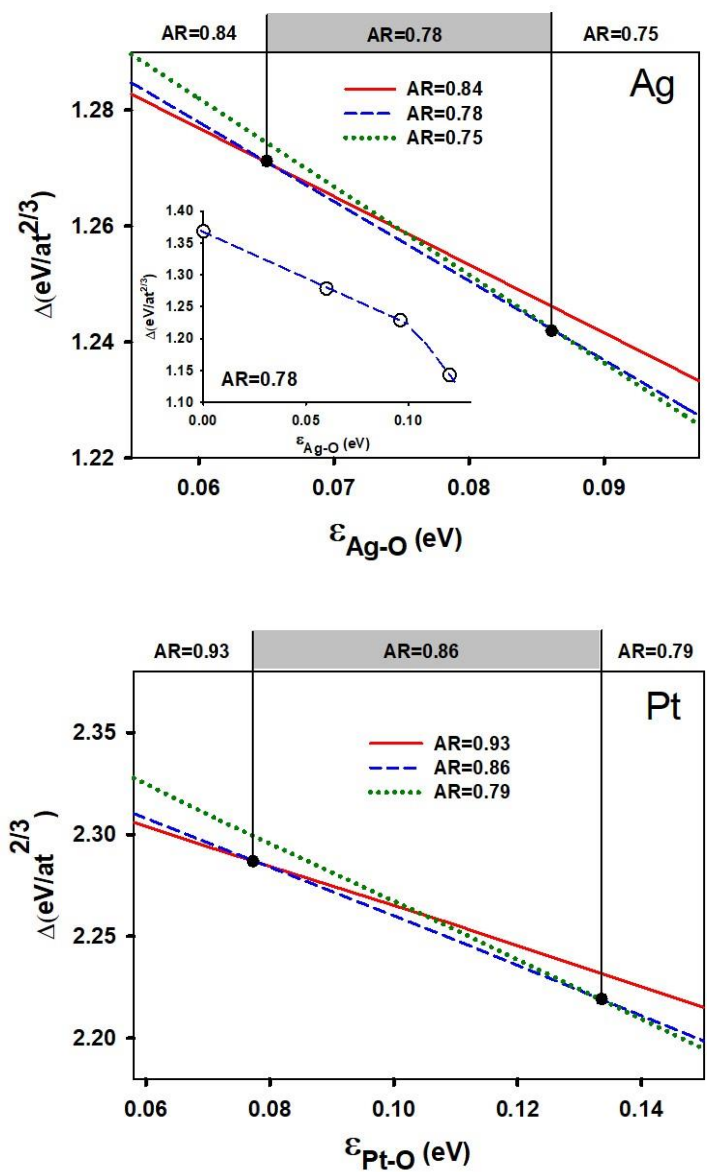

Fig 10: Excess energy $\Delta$ of metallic NPs deposited on the amorphous silica support as a function of the free parameter $\varepsilon_{M-O}$. Insert: Ag NP of AR = 0.78, showing a linear regime for weak interactions (≤ 0.1 eV). Main panels: enlargement of the $\Delta$ vs $\varepsilon_{M-O}$ curves for the three AR given in the figure. Greyed regions: range of $\varepsilon_{M-O}$ compatible with the experimental values AR = 0.78 for Ag and 0.84 for Pt.

Figure 11

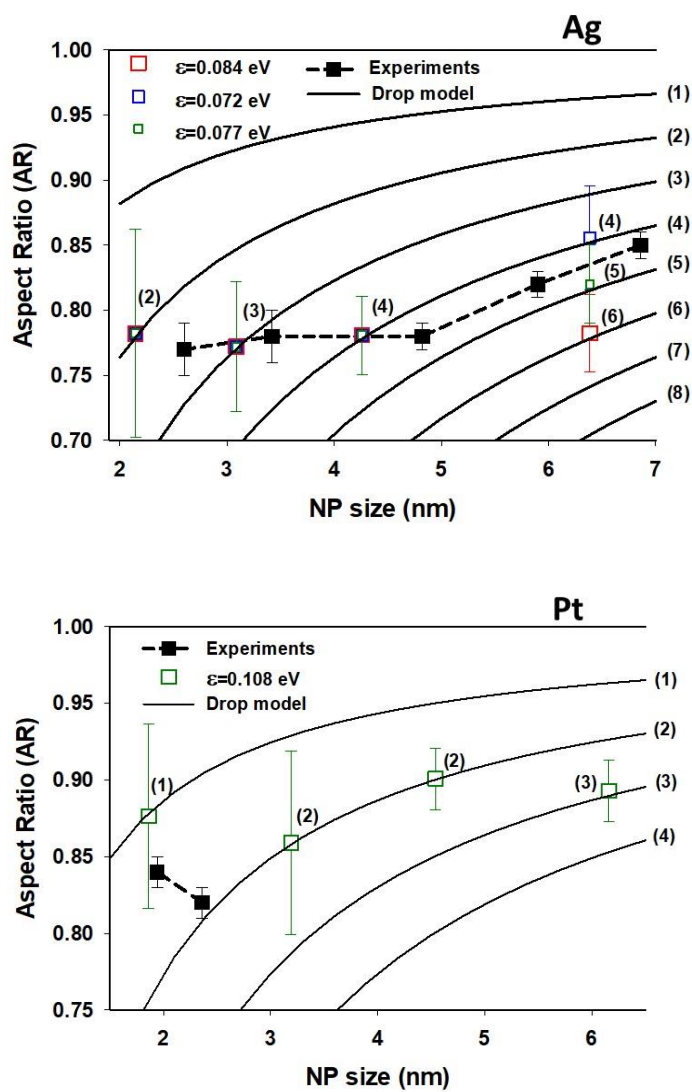

Fig 11: Empty squares: most stable aspect ratio as a function of the NP size as determined by minimization of the excess energy (Eq. 4); upper panel: Ag NP, with 3 values of the Ag-silica interaction parameter $\varepsilon_{Ag-O}$ given in the figure; lower panel: Pt NP with $\varepsilon_{Pt-O}$ = 108 meV. Between parentheses is the number of removed layers at the interface with the amorphous silica support. Solid squares: experimental data obtained by GISAXS. Dashed lines: guide to the eye through the data. Solid lines: simple theoretical model (homogeneous truncated drop) with AR = 1 - $id/D$ where $D$ is the size (diameter), $d$ is the interlayer distance (0.236 nm for Ag and 0.226 nm for Pt) and $i$ is the number of removed layers at the bottom (given between parentheses on the right-hand side of the figures).

Figure 12

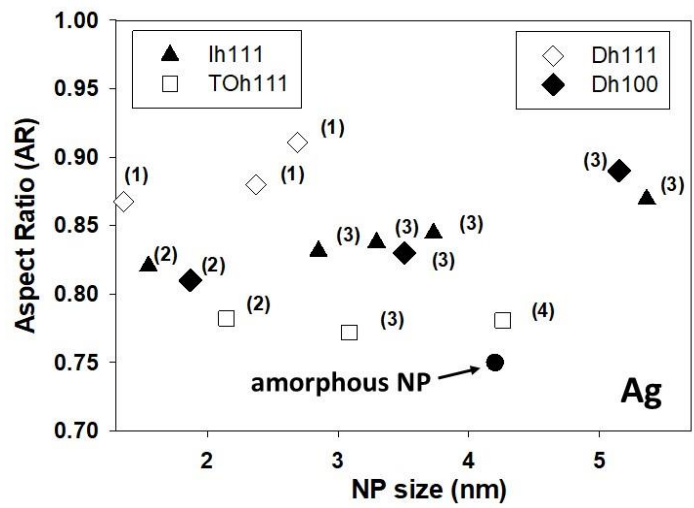

Fig. 12: Most stable aspect ratio as a function of the Ag NP size as determined by minimization of the excess energy (Eq. 4) for $\varepsilon_{Ag-O}$ = 77 meV, and for different structures (Ih, Dh and TOh) deposited on the facet given in the figure.